\documentclass[paper,preprint]{jpsj2}
\usepackage[]{graphicx,color}
\usepackage{bm}

\title{%
Frustration-induced dodecamer ordering \\
in the double-exchange spin ice model on the kagom\'e lattice
}

\author{%
Yoshihiro {\sc Shimomura}\thanks{E-mail address: shimomura@phys.aoyama.ac.jp},
Shin {\sc Miyahara} and Nobuo {\sc Furukawa} 
}

\inst{%
Department of Physics and Mathematics, 
Aoyama Gakuin University, 
Sagamihara 229-8558, Japan
}

\recdate{\today}

\abst{%

We investigate a detail of a dodecamer cluster ordering
in a double-exchange spin ice model on a kagom\'e lattice.
In frustrated systems, ordinary spin orderings are suppressed
and macroscopic degeneracy remains down to low temperatures.
In some frustrated systems, the degeneracy is
lifted due to residual interactions and cluster orderings 
are stabilized.
In the present model, the spin ice state is first formed
at intermediate temperatures, and further entropies are
released at lower temperatures as the dodecamer phase emerges.
Since the spin symmetry is not broken in the dodecamer phase,
there still exists macroscopic degeneracy.
At further low temperatures,
a possible spin ordering due to inter-dodecamer interactions
is proposed.
We discuss that such a multiple-site clustering larger than
a bond-pair might be generic to frustrated systems
where macroscopic degeneracy is lifted by residual
interactions.

}

\kword{%
dodecamer, cluster order, double-exchange model, kagom\'e lattice, 
frustration, metal, spin ice, Monte Carlo simulation
}

\begin{document}
\maketitle

\section{Introduction}

Intensive studies have been devoted to frustrated systems 
because of their unexpected behaviors.
In spin systems, a frustration arises 
when all interaction energies between localized spins 
can not be minimized simultaneously in the classical picture. 
It is known that the frustration suppresses the long-range order 
and often induces the novel ground state. 
The macroscopically degenerate ground state 
accompanied with a residual entropy 
is one of the well-known states 
in frustrated Ising spin systems. 
Such a state is seen in the antiferromagnetic (AF) Ising model 
on a triangular\cite{Wannier_50}, 
a kagom\'e\cite{Syozi_51, Kano_53} and 
a pyrochlore\cite{Anderson_56} lattices. 
On the other hand, 
a peculiar cluster ordered state can be induced by the frustration. 
In the Majundar-Ghosh model\cite{Majumdar_69} 
and the Shastry-Sutherland model\cite{Shastry_81} 
which are frustrated quantum spin systems, 
the dimer-singlet states with a spin gap 
are realized as the unique ground states. 
The entropies in the systems are released 
by the dimer-singlet formations for all localized spins, 
which can be viewed as a cluster ordering 
since the systems are tiled by the dimer-singlets. 
Such dimer-singlet ground states are driven by 
geometrical frustrations and quantum effects of the interactions. 
The cluster ordering can be the way to stabilize the states 
in the frustrated systems.

Actually, such a cluster order is seen even in realistic materials.
For example, the Shastry-Sutherland model 
is realized in the two-dimensional spin-1/2 orthogonal dimer system 
${\rm SrCu_{2}(BO_{3})_{2}}$, 
where Cu sites are tiled by the dimer-singlet. 
This material is novel in terms of the existence of 
the exact dimer-singlet ground state with the finite spin gap
in the two-dimensional system\cite{Kageyama_99,Miyahara_99}.  
Furthermore, as possible other experimental examples 
of the spin cluster order in frustrated spin systems,
hexamers of ${\rm Cr^{3+}}$ 
in a cubic spinel ${\rm ZnCr_{2}O_{4}}$\cite{Lee_02} 
and octamers of ${\rm Ir^{3+}}$ and ${\rm Ir^{4+}}$ 
in a thiospinel ${\rm CuIr_{2}S_{4}}$\cite{Radaelli} 
have been reported.

\begin{figure}
\begin{center}
\includegraphics[width=12cm,keepaspectratio]{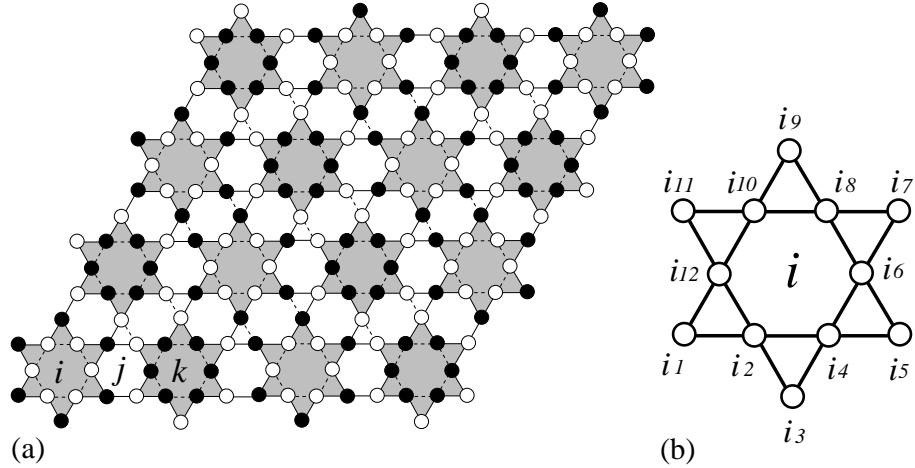}
\end{center}
\caption{(a) Dodecamer phase. 
Open and closed circles represent out- and in-spins 
for the up-triangles, respectively. 
The dodecamer is represented by the shadowed spin cluster. 
Each dodecamer is labeled $i$, $j$, $k$ and others. 
(b) Twelve sites which construct the $i$-th dodecamer.}
\label{fig:ddcmr_phase}
\end{figure}

In this way, cluster orders in frustrated spin systems are 
seen in both theoretical and experimental researches, 
and have attracted much attentions. 
Even in electron systems, 
similar cluster ordering induced by the frustration can be realized.
As an example of such cluster orders in electron systems, 
we have reported dodecagonal localized spin cluster, 
``dodecamer'', order in a double-exchange spin ice (DESI) model 
on the kagom\'e lattice\cite{Shimo_04} 
[see Fig. \ref{fig:ddcmr_phase}(a)]. 
In the DESI model, localized spins have local uniaxial anisotropies 
for triangles which constitute the kagom\'e lattice: 
Localized spins are forced to point in the directions 
connecting the center of mass and the three vertexes of the triangles, 
that is, they point either inward (in-spin) or outward (out-spin) 
for up-triangles.  
There is an effective ferromagnetic interaction between localized spins 
due to the double-exchange (DE) mechanism\cite{Zener_51}. 
The situation that there are 
the ferromagnetic interactions and the uniaxial anisotropies 
for localized spins 
is similar to that in the spin ice systems\cite{Bramwell_01_science}. 
From an analogy with the spin ice systems, 
it is obvious that 
the DESI model has the frustration and a spin-ice-like behavior. 
The dodecamer formation is driven by the frustration and 
the kinetic energy gain due to the DE mechanism: 
At sufficiently low temperatures, 
electrons move along a certain selected path 
to gain the kinetic energy under the frustration. 
The dodecamer formation is a consequence 
of the path selection by electrons in order to gain the kinetic energy. 
In the dodecamer phase, 
the kagom\'e lattice can be completely tiled by dodecamers, 
which leads to the dodecamer order with the transilational symmetry breaking. 
Note that there are two kinds of dodecamers: 
One is the dodecamer with in-spins on the tips 
$i_{2n-1}$ ($n=1,2, \cdots ,6$) shown in Fig. \ref{fig:ddcmr_phase}(b) 
and the other is out-spins. 
From the existence of two dodecamers, 
it is found that the spin symmetry is not broken in the dodecamer phase.

In this paper, we focus on the behaviors of the entropy release 
in the DESI system.  
As the temperature is lowered, 
the thermodynamics in the DESI system is as follows. 
(1) The spin-ice-like state is stabilized at intermediate temperatures. 
(2) The dodecamer order appears at sufficiently low temperatures. 
In the spin-ice-like state, there still remain 
a large part of degrees of freedom for localized spins, i.e., 
there is a finite entropy which is almost consistent with 
the residual entropy of the AF Ising model on the kagom\'e lattice
\cite{Kano_53}. 
A large part of the entropy is released by the dodecamer ordering. 
There is a possibility 
that the entropy still remains even in the dodecamer phase, 
since each dodecamer is two-fold degenerate. 

\section{Double-Exchange Spin Ice Model}
In this section, let us introduce 
the DESI model on the kagom\'e lattice, 
which is the Anderson-Hasegawa model\cite{Anderson-Hasegawa_55} 
with local uniaxial anisotropies. 
The mechanism for the frustration in this system 
is similar to that in the spin ice systems\cite{Bramwell_01_science}. 
It is expected that 
similar degeneracy as in the spin ice systems exists 
in the DESI model. 

\subsection{Anderson-Hasegawa model}

The Anderson-Hasegawa model\cite{Anderson-Hasegawa_55} 
is a modified DE model with infinitely large Hund's-rule couplings: 
\begin{equation}
\hat{H} = -\sum_{\langle i,j \rangle} 
t( \bm{S}_{i}, \bm{S}_{j} )
\left( c_{i}^{\dagger} c_{j} + h.c.\right)
- \mu \sum_{i} c_{i}^{\dagger} c_{i},
\label{eq:EffDEmodel}
\end{equation}
where $c_{i}^{\dagger}(c_{i})$ is 
an electron operator which creates (annihilates) an electron 
with a spin parallel to a localized spin at site $i$, ${\bm S}_i$, 
and $\mu$ is the chemical potential in the grand canonical ensemble. 
Here, $t( \bm{S}_{i}, \bm{S}_{j})$ is given in the form, 
\begin{equation}
t( \bm{S}_i, \bm{S}_j ) = 
t \cos \left( \frac{\theta_{ij}}{2} \right), 
\label{eq:t_eff}
\end{equation}
where $t$ is a transfer integral and $\theta_{ij}$ is a relative angle 
between the localized spins $\bm{S}_{i}$ and $\bm{S}_{j}$. 
The angle $\theta_{ij}$ is given by 
\begin{equation}
\cos \theta_{ij} = \cos \theta_{i} \cos \theta_{j} 
+ \sin \theta_{i} \sin \theta_{j} \cos ( \phi_{i} - \phi_{j}),
\end{equation}
where $(\theta_{i},\phi_{i})$ is an angle of ${\bm S}_i$ 
in the polar coordinate. 
Note that $t(\bm{S}_i, \bm{S}_j)$ has a maximum value $t$ 
when $\bm{S}_i$ and $\bm{S}_j$ are parallel. 
Thus, ferromagnetic alignment of the localized spins broadens 
the bandwidth of the system, which produces a kinetic energy gain. 
Therefore, there is an effective ferromagnetic interaction 
between the nearest-neighbor (n.n.) localized spins. 
We set the transfer integral $t \equiv 1$, hereafter. 

\subsection{Frustrated electron system : double-exchange spin ice model}

\begin{figure}
\begin{center}
\includegraphics[width=13cm,keepaspectratio]{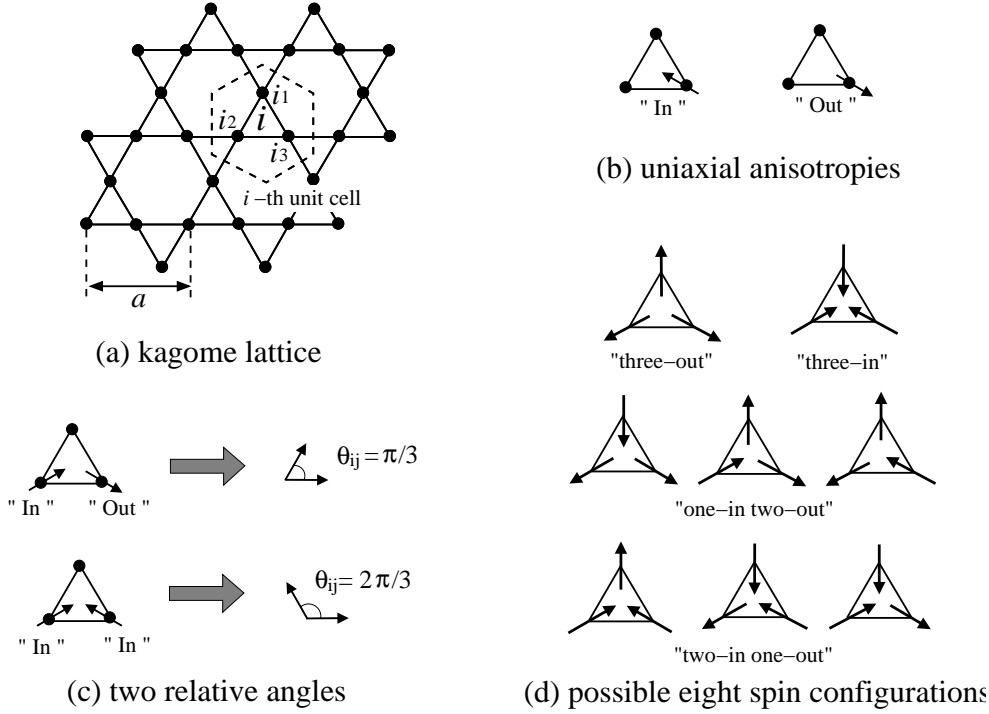}
\end{center}
\caption{
(a) Kagom\'e lattice. 
The $i$-th unit cell surrounded by the dashed line 
consists of three sites $i_{1}$, $i_{2}$ and $i_{3}$. 
$a$ is a lattice constant.
(b) Uniaxial anisotropies for localized spins. 
Localized spins have two degrees of freedom, 
``in'' and ``out'', for a triangle. 
(c) two relative angles in the DESI system:
The relative angle $\theta_{ij}$ between in-spin and out-spin, 
and between the same spins, 
are $\frac{\pi}{3}$ and $\frac{2\pi}{3}$, respectively. 
(d) Possible eight configurations for an up-triangle in the kagom\'e lattice. 
} 
\label{fig:uniaxial}
\end{figure}

We consider the Anderson-Hasegawa model 
on the kagom\'e lattice [see Fig. \ref{fig:uniaxial}(a)], and 
introduce uniaxial anisotropies for the localized spins as follows:  
They are forced to be either in-spin or out-spin 
for up-triangles on the kagom\'e lattice 
[see Fig. \ref{fig:uniaxial}(b)]. 
Under the conditions, relative angles between the localized spins, 
$\theta_{ij}$ (Eq.(\ref{eq:t_eff})), can take only two values, i.e.,  
$\theta_{ij}^1 \equiv \pi/3$ 
(the angle between an in-spin and an out-spin) and 
$\theta_{ij}^2 \equiv 2 \pi/3$ (the angle between the same spins)
[see Fig. \ref{fig:uniaxial}(c)]: 
The transfer integral of electrons $t(\bm{S}_i,\bm{S}_j)$ has 
$t_1 \equiv \cos (\theta_{ij}^1/2) = \sqrt{3}/2$ and 
$t_2 \equiv \cos (\theta_{ij}^2/2) = 1/2$. 
Thus, electrons can easily move through the ferromagnetic bonds 
with the transfer integral $t_1$ 
due to the kinetic energy gain.

To clarify the frustration in the DESI system, 
we introduce a pseudospin representation. 
We define a pseudospin corresponding to ${\bm S}_i$, 
\begin{equation}
\tau_{i} = 
\begin{cases}
+1 & \text{($\bm{S}_{i}$ is an out-spin)}, \\
-1 & \text{($\bm{S}_{i}$ is an in-spin)}.
\end{cases}
\end{equation}
In this pseudospin picture, 
the Hamiltonian (\ref{eq:EffDEmodel}) is rewritten in the form 
\begin{align}
\hat{H} = \sum_{\langle i,j \rangle} t \left( \tau_{i}, \tau_{j} \right)
\left( c_{i}^{\dagger} c_{j} + h.c. \right)
- \mu \sum_{i} c_{i}^{\dagger} c_{i},
\label{eq:DESImodel}
\end{align}
where 
\begin{align}
t \left( \tau_{i}, \tau_{j} \right) = 
\begin{cases}
- t_{1} & \text{($\tau_{i}=-\tau_{j}$)}, \\
- t_{2} & \text{($\tau_{i}=\tau_{j}$)}.
\end{cases}
\end{align}
The effective ferromagnetic interaction 
between the n.n. localized spins is regarded as 
the effective AF interaction between the n.n. pseudospins. 
Thus, there is the n.n. AF interaction on the kagom\'e lattice.
Using the pseudospin picture, 
it is easily understood that the system has the frustration.

\subsection{Spin ice state in the double-exchange spin ice model}

In the DESI system, 
each triangle has four types of spin configurations, i.e., 
``three-in'', ``three-out'', ``two-in one-out'' and ``one-in two-out'' 
as shown in Fig. \ref{fig:uniaxial}(d). 
These correspond to ``three-down'', ``three-up'', 
``two-down one-up'' and ``one-down two-up'', respectively, 
in the pseudospin picture. 
Naturally, as the temperature is lowered, 
three-in and three-out configurations are unfavored 
by the ferromagnetic interaction due to the DE mechanism. 
In that case, the majority of localized spin configurations on the triangles 
are two-in one-out and one-in two-out. 
Here, each configuration is three-fold degenerate. 
A total spin state in the system is expected to 
have the macroscopic degeneracy 
since there still exist many degrees of freedom 
that each triangle satisfies either two-in one-out or one-in two-out. 
From the analogy with the spin ice systems\cite{Bramwell_01_science}, 
an ``ice state'' in the DESI system is defined as a state with 
the macroscopic degeneracy and the two types of spin configurations, 
i.e., two-in one-out and one-in two-out.

We can characterize the ice state 
by using a vector spin chirality $\bm{v}_{i}$ defined by
\begin{equation}
\bm{v}_{i} = -\frac{2}{\sqrt{3}}\left(
  \bm{S}_{i_{1}}\times \bm{S}_{i_{2}} 
+ \bm{S}_{i_{2}}\times \bm{S}_{i_{3}}
+ \bm{S}_{i_{3}}\times \bm{S}_{i_{1}} \right), 
\label{eq:chiral}
\end{equation}
where $i_{1}$, $i_{2}$ and $i_{3}$ represent three sites of the triangles 
in the kagom\'e lattice shown in Fig. \ref{fig:uniaxial}(a). 
Here, we set $|\bm{S}_{i_{\alpha}}| \equiv 1$ ($\alpha$ = 1, 2 and 3). 
$\bm{v}_{i}$ has only $z$-component 
since localized spins in the DESI system are coplanar, 
where we consider that the kagom\'e lattice is spread in $xy$-plane. 
$v_{i}^{z}$ has -3 (three-in and three-out) 
or 1 (one-in two-out and two-in one-out). 
It is expected that the thermal average of the vector spin chirality 
$\langle v_{i}^{z} \rangle$ vanishes 
if the four types of configurations are realized equally. 
On the other hand, $\langle v_{i}^{z} \rangle$ approaches unity 
when a large number of spin configurations for triangles are 
two-in one-out and one-in two-out. 
Furthermore, we define a fluctuation of $v_{i}^{z}$, 
which is given in the form 
\begin{equation}
\chi_{i}^{z} = \frac{\langle (v_{i}^{z})^2 \rangle 
- \langle v_{i}^{z} \rangle ^2}{T}.
\end{equation}
To see the existence of the ice state, 
we calculated the total vector chirality 
$v_{z} = \sum_{i} \langle v_{i}^{z} \rangle / N_{\rm T}$ 
and the total fluctuation $\chi_{z}= \sum_{i}\chi_{i}^{z}/ N_{\rm T}$. 
Here $N_{\rm T}$ represents the number of triangles on the kagom\'e lattice 
which is given by $\frac{2}{3}N_{\rm S}$, 
where $N_{\rm S}$ is the number of sites. 
The summation runs over all up- and down-triangles on the kagom\'e lattice.

One may think that the DESI model on the kagom\'e lattice 
could be completely mapped to the n.n. AF Ising model on the same lattice, 
which has a macroscopically degenerate ground state
\cite{Syozi_51,Kano_53}. 
However, the interaction of the DESI system are determined 
by the kinetics of electrons, 
and it is expected that the behavior of the DESI model is
different from that of the frustrated Ising spin system 
with the n.n. interaction.

\section{Results : Thermodynamics of the Double-Exchange Spin Ice System on the Kagom\'e Lattice}

In this section, we present the Monte Carlo (MC) results of 
the DESI model on the kagom\'e lattice at finite temperatures. 
We applied a MC technique 
to the calculation of thermodynamic quantities 
by using the grand canonical ensemble [see Appendix].
We apply the Metropolis algorithm 
for the updates of spin configurations and 
typically run 100,000 MC steps for measurement 
after 10,000 thermalization steps.

In ref. 11, we have already reported the dodecamer order 
of localized spins at sufficiently low temperatures.  
The dodecamer order is realized in a wide range of $\mu$, 
which is roughly estimated as $-0.3 \lesssim \mu \lesssim 0.2$. 
Thus, the DESI models around $\mu \sim 0$ 
are expected to show qualitatively similar behaviors. 
Therefore, we treat the DESI model at $\mu =0$, hereafter.

\begin{figure}
\begin{center}
\includegraphics[width=15cm,keepaspectratio]{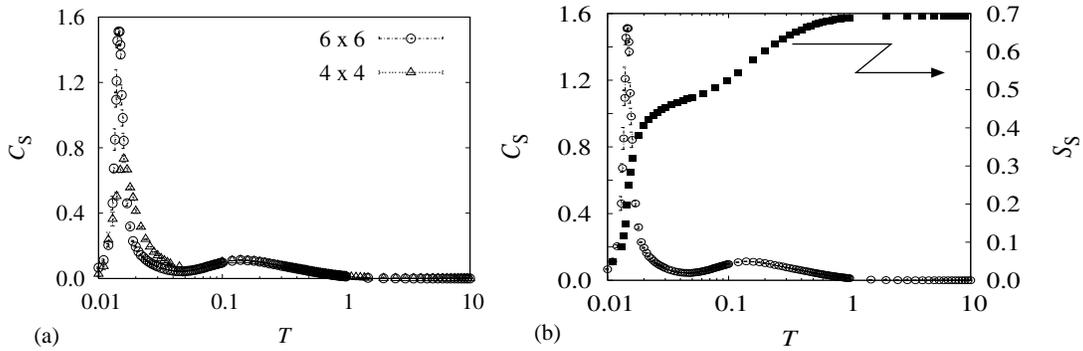}
\end{center}
\caption{(a) Temperature dependence of the specific heat, 
$C_{\rm S}$, for the system 
with 4 $\times$ 4 (= 48 sites) $(\bigtriangleup)$ 
and 6 $\times$ 6 (= 108 sites) $(\Box)$ unit cells at $\mu = 0$. 
The specific heat has a double-peak structure, 
a broad peak ($T \sim 0.2$) and 
a sharp peak ($T \sim 0.015$). 
(b) Temperature dependence of 
the specific heat, $C_{\rm S}$, and the entropy, 
$S_{\rm S}$, ($\blacksquare$)
for the system with 6 $\times$ 6 unit cells. 
}
\label{fig:spe}
\end{figure}
Let us investigate the behaviors on the DESI model at finite temperatures. 
In order to see how the entropy of localized spins 
is released in the system, 
we have investigated a specific heat and localized spin configurations. 
As explained in Appendix, 
the specific heat of the system can be decomposed into 
an electronic part $C_{\rm E}(T)$ and 
a localized spin part $C_{\rm S}(T)$. 
The information of localized spin configurations 
is included in $C_{\rm S}(T)$. 
In particular, we are interested in the behaviors of the localized spins 
at low temperatures. 
Thus, we only focus on $C_{\rm S}(T)$, 
which is defined as the specific heat per site hereafter. 
In Fig. \ref{fig:spe}(a), we show 
the temperature dependences of $C_{\rm S}(T)$, 
for the system with $4 \times 4$ unit cells ($N_{\rm S}=48$) 
and $6 \times 6$ unit cells ($N_{\rm S}=108$) 
at $\mu = 0$. 
$C_{\rm S}(T)$ has a double-peak structure: 
A broad peak ($T \sim 0.2$) and 
a sharp one ($T \sim 0.015$). 
In Fig. \ref{fig:spe}(b), 
the temperature dependence of the $C_{\rm S}$ 
and the entropy $S_{\rm S}$ obtained from $C_{\rm S}$ 
for the system with 6 $\times$ 6 unit cells are shown. 
These results provide us a picture 
that there exist two steps 
to release the entropy in the DESI system.

\subsection{Crossover to the ice state}

Let us consider the broad peak in $C_{\rm S}(T)$
at $ T \sim 0.2$.
As mentioned in Sec. 2.3,
it is expected that the system exhibits an ice state
as temperature is lowered.
We speculate that the broad peak in $C_{\rm S}(T)$
is associated with the formation
of the ice state.
In order to clarify this idea,
we investigate the entropy per site $S_{\rm S}(T)$ from $C_{\rm S}(T)$,
\begin{align}
S_{\rm S}(T) &= \log2
- \int_{T}^{\infty}
\frac{C_{\rm S}(T^{\prime})}{T^{\prime}} dT^{\prime}.
\label{eq:entropy}
\end{align}
Here we use $S_{\rm S}(\infty)= \log2$,
since all spin configurations
with the number of states $2^{N_{\rm S}}$ can be realized
in the high temperature limit ($T=\infty$).
The data for  $S_{\rm S}(T)$
are shown in Fig. \ref{fig:spe}(b).

In the high temperature region, $S_{\rm S}(T)$ deviates from $\log 2$
at $T \lesssim 1$, since the interactions are
mediated by the DE mechanism.
In the intermediate temperature region,
  we see that $S_{\rm S}(T)$ has a
plateau structure at $T \sim 0.05$ with
$S_{\rm S}(T) \simeq 0.5$.
Since the spin ice system on a kagom\'e lattice can be
mapped to an AF Ising model on the identical lattice,
the entropy of the ice state for the DESI system
is roughly estimated from the residual entropy
of the AF Ising model on the kagom\'e lattice\cite{Kano_53},
$S_{\rm Ising} \simeq 0.5$, which is consistent with
the observed value for $S_{\rm S}$.

\begin{figure}
\begin{center}
\includegraphics[width=8cm,keepaspectratio]{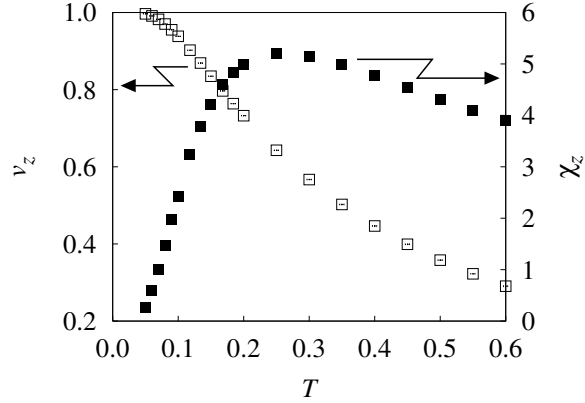}
\end{center}
\caption{ Temperature dependence of the vector spin chirality ${v}_z (\Box)$
and the susceptibility of the chirality $\chi_z (\blacksquare)$ 
for the system with $6 \times 6$ unit cells at $\mu = 0$. 
$v_{z}$ approaches unity as the temperature is lowered. 
$\chi_{z}$ has a broad peak at $T \sim 0.25$.}
\label{fig:chiral}
\end{figure}

An alternative clarification is given by the temperature dependence of
$v_{z}$ and $\chi_{z}$.
The results for the system
with 6 $\times$ 6 unit cells (108 sites) and $\mu = 0$
are shown in Fig. \ref{fig:chiral}.
$\chi_z$ has the maximum value at $T \sim 0.2$,
and then, below which it rapidly decreases until $T \sim 0.05$.
Similarly, it is found that
$v_z$ has a inflection point
at around $T \sim 0.2$,
and saturates below $T \sim 0.05$.

These results suggest that
the broad peak in $C_{\rm S}(T)$
at $ T \sim 0.2$  is due to a crossover to the ice state.
At $T\sim 0.05$, the ice state is
stabilized,
where a macroscopic
number of thermally degenerate states still remain. From the
analogy with the spin ice system\cite{Bramwell_01_science},
it is considered that the origin of the ice state in the DESI system
is the short-range (n.n.) interaction between localized spins.
The ice state in which a large number of the entropy remains
is indicated by the broad peak in the specific heat and
is driven by the short-range ferromagnetic interaction.

\subsection{Transition to the dodecamer ordered state}

In this section, we focus on the sharp peak in the specific heat. 
The sharp peak means further entropy release in the system 
which results from the reduction of degrees of freedom for localized spins. 
As an origin of such an entropy release, 
it is considered that some kind of ordering of spins are generated 
due to the residual effects in addition to the short-range interaction 
in the model. 
An effective long-range interaction 
due to the kinetic energy gain of electrons 
is one candidate for such effects, 
since higher order hopping process of electrons 
is considered to be important at low temperatures. 
Indeed, the dodecamer order 
driven by the kinetic energy gain of electrons 
appears at low temperatures in the system\cite{Shimo_04}. 
Thus, it is expected that the sharp peak corresponds to a change 
from the ice state to the dodecamer ordered state.

The order parameter for the dodecamer formation 
is characterized by using bond correlations, and  
is defined by 
\begin{equation}
d_{i} \equiv \frac{1}{18} 
\left( 
- \sum_{n=1}^{12} \tau_{i_{n}} \tau_{i_{n+1}}
+ \sum_{m=1}^{6} \tau_{i_{2m}} \tau_{i_{2m+2}}
\right),
\label{eq:order_param}
\end{equation}
where a site $i_{n}$ is indicated in Fig. \ref{fig:ddcmr_phase}(b) and 
we set $i_{13} \equiv i_{1}, \ i_{14} \equiv i_{2}$. 
Note that the order parameter $d_{i}$ is a unity 
when the dodecamer is formed. 
The dodecamer structure factor is given in a form 
\begin{equation}
D_{\bm q} = \frac{1}{N_{\rm C}} \sum_{i,j} 
	e^{i \bm{q} \cdot (\bm{r}_{i}- \bm{r}_{j})}
	\langle d_{i} d_{j} \rangle ,
\label{eq:Dq}
\end{equation}
where $N_{\rm C}$ is the number of unit cells 
($N_{\rm C} \equiv \frac{1}{3}N_{\rm S}$) and 
$\sum_{i,j}$ represents the summation 
for any pairs of the order parameter 
shown in Fig. \ref{fig:ddcmr_phase}(a).
$D_{\bm q}$ has three independent maximum peaks at 
$\bm{Q}_{1}={}^t(0,2 \pi / \sqrt{3} a)$, 
$\bm{Q}_{2}={}^t( \pi / a, \pi / \sqrt{3} a)$ 
and $\bm{Q}_{3}={}^t( \pi / a, - \pi / \sqrt{3} a)$, 
those characterize the dodecamer order. 
An average of maximum peaks, 
$D_Q \equiv \{ D_{\bm{Q}_1}+D_{\bm{Q}_2}+D_{\bm{Q}_3} \}/3$, 
diverges increasing the system size 
at sufficiently low temperatures\cite{Shimo_04}. 
This result indicates that 
the dodecamer order is realized in the DESI system on the kagom\'e lattice. 
Note that the change from the ice state 
to the dodecamer ordered state is a phase transition 
since the translational symmetry is broken due to the dodecamer ordering.

\begin{figure}
\begin{center}
\includegraphics[width=80mm,keepaspectratio]{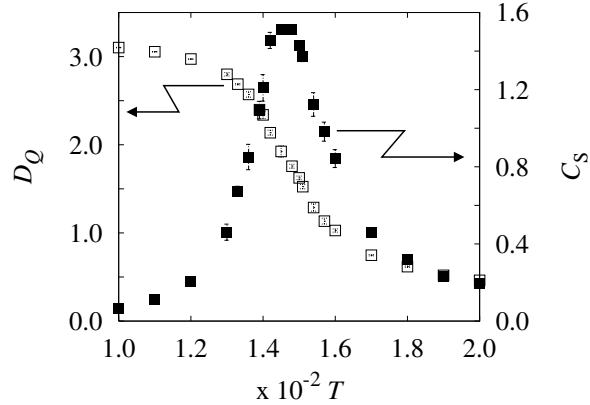}
\end{center}
\caption{Temperature dependence of $D_Q$ 
and specific heat $C_{\rm S}$ for the system with $6 \times 6$ unit cells 
and $\mu = 0$.}
\label{fig:Dq_and_spe}
\end{figure}

Temperature dependences of $D_{Q}$ and specific heat $C_{\rm S}$
for $N_{\rm C} = 6 \times 6$ and $\mu = 0$ 
are also shown in Fig. \ref{fig:Dq_and_spe}. 
$D_Q$ increases rapidly 
at around the sharp peak temperature in the specific heat. 
The temperature at the sharp peak 
is almost consistent with the inflection point of $D_Q$. 
Thus, we conclude 
that the sharp peak corresponds to a transition to the dodecamer order
driven by the effective long-range interaction 
due to the kinetic energy gain. 
In fact, the majority of the entropy release in the DESI system 
is caused by the dodecamer order as shown in Fig. \ref{fig:spe}(b). 
The similar behavior is observed in the system with $4 \times 4$ unit cells. 
Note that the sharp peak is smeared out by the finite-size effect.

\subsection{Stability of the dodecamer ordered state}

A stability and an origin for the dodecamer ordered state 
have been discussed in Ref. 11. 
Let us now present brief summary. 
(1) The dodecamer order is realized in a wide doping region 
$1/3 \lesssim n \lesssim 1/2 \ (-0.3 \lesssim \mu \lesssim 0.2)$, 
where $n$ is the number of particles per site. 
(2) The dodecamer order is consequence of a selection 
of a certain path, along which electrons can move easily, i.e., 
the kinetic energy gain. 
The result of (2) is concluded by the density of states: 
An energy gap is not opened by the dodecamer ordering. 
From this result, we concluded that the dodecamer ordered state is metallic. 
The result of (1) also means that the dodecamer order is not sensitive 
to the Fermi surface. 
In this way, the results of (1) and (2) are consistent with each other. 
Considering the DE mechanism which gains the kinetic energy 
by broadening the bandwidth, 
the weak $n$-dependence of the dodecamer ordered state 
is understood as follows. 
The ferromagnetic interaction is dominant among the localized spins 
at around $\mu=0$ where the kinetic energy is mostly gained. 
Although there is the spin structure 
that is characterized by the finite wave vectors, 
the spin structure is not originated from the nesting of 
the Fermi surface, but from the frustration, i.e., 
the uniaxial anisotropy.

\begin{figure}
\begin{center}
\includegraphics[width=8cm]{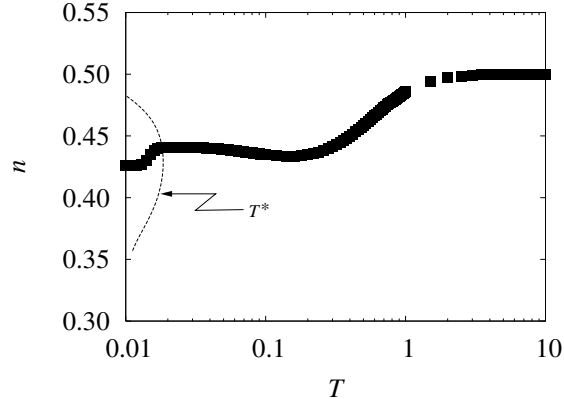}
\end{center}
\caption{
Temperature dependence of $n$ for the system 
with $N_{\rm C}=6 \times 6$ and $\mu =0$, 
where black squares represent $n$ at each temperature. 
The dashed line represents the approximate phase boundary 
for the dodecamer ordered state $T^{*}$ estimated from $D_{Q}$ 
[see Ref. 11 for further details]. 
There is the two-step structure 
which corresponds to the crossover to the ice state 
and the transition to the dodecamer ordered state as well as the entropy. 
}
\label{fig:N_vs_T}
\end{figure}

We show the temperature dependence of $n$ for the system 
with $N_{\rm C} = 6 \times 6$ and $\mu=0$ in Fig. \ref{fig:N_vs_T}. 
The two-step behavior can be found, which 
corresponds to the crossover to the ice state and 
the phase transition to the dodecamer ordered state. 
$n$ decreases gradually from $T \sim 1$ to 0.2. 
In the ice state, $n$ keeps the almost same values 
which can be regarded as a plateau structure.  
Going down to further low temperatures, 
$n$ shows a rapid reduction at around the sharp peak of $C_{\rm S}$. 
The behavior of $n$ is consistent with that of the entropy. 
The values of $n$ at $\mu = 0$ are also included in 
the region $\frac{1}{3} \lesssim n \lesssim \frac{1}{2}$. 
It is expected that the temperature dependence of $n$  
is not affected drastically by changing $\mu$ around zero. 
Thus, similar arguments are considered to be useful 
in the region of $\mu$ 
where the dodecamer ordered state is observed.

\subsection{Possibility of further transition in the dodecamer ordered phase}

\begin{figure}
\begin{center}
\includegraphics[width=8cm,keepaspectratio]{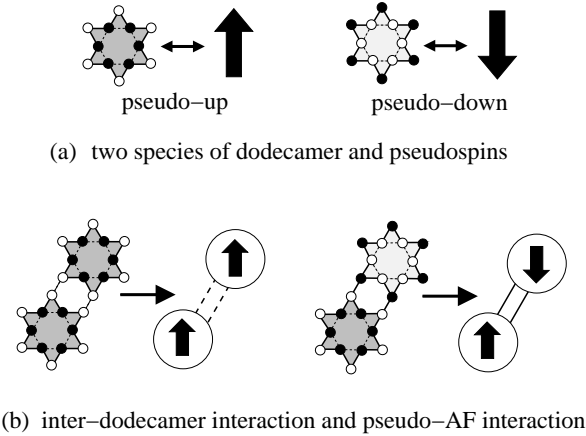}
\end{center}
\caption{(a) Two species of dodecamer: pseudo-up and pseudo-down. 
A pseudo-up and a pseudo-down have out-spins and in-spins 
on pointed tips of the dodecamer, respectively. 
(b) Inter-dodecamer interaction:  
An inter-dodecamer interaction behaves like the AF interaction, 
because there are two bonds with the transfer integral $t_{1}$ 
between a pseudo-up and a pseudo-down. 
On the other hand, there are two bonds with $t_{2}$ 
between the same pseudo-spins. 
Solid and dashed double lines represent pseudo-AF and 
pseudo-ferromagnetic bond, respectively. 
}
\label{fig:inter_ddcmr}
\end{figure}

In the dodecamer phase, e.g., Fig. \ref{fig:ddcmr_phase}(a), 
we can find that dodecamers are two-fold degenerate. 
One has out-spins on the tips, and the other has in-spins 
as in Fig. \ref{fig:inter_ddcmr}(a). 
Two dodecamers can be regarded as 
``pseudo-up'' and ``pseudo-down'', respectively. 
Then the dodecamer order can be described using pseudospins 
on a triangular lattice. 
Due to the DE mechanism, 
a pseudo-up and a pseudo-down are connected by the two bonds 
with the transfer integral $t_{1}$ [see Fig. \ref{fig:inter_ddcmr}(b)(right)], 
while the same pseudospins are connected by the two bonds 
with the transfer integral $t_{2}$ [see Fig. \ref{fig:inter_ddcmr}(b)(left)]. 
There is an AF interaction between the pseudospins 
since $t_{1} > t_{2}$. 
Thus, in the dodecamer phase, 
the problem of the DESI system on the kagom\'e lattice 
can be mapped onto that of psuedospin systems 
with the effective AF interaction on the triangular lattice. 
When the random pseudospin configurations are realized,  
it is expected that the entropy $S_{\rm S}(T)$ 
is $\frac{1}{12}\log2 \simeq 0.06$ 
since the dodecamer has two degrees of freedom. 
However, it is considered that 
some of the entropy $S_{\rm S}(T)$ per site are released 
due to the AF interaction. 
For example, in the AF Ising model on the triangular lattice, 
the exact residual entropy is known to be about 0.34 per site 
\cite{Wannier_50}, 
which corresponds to the residual entropy of $0.34/12 \simeq 0.03$ 
per site for the present model. 
\begin{figure}
\begin{center}
\includegraphics[width=80mm,keepaspectratio]{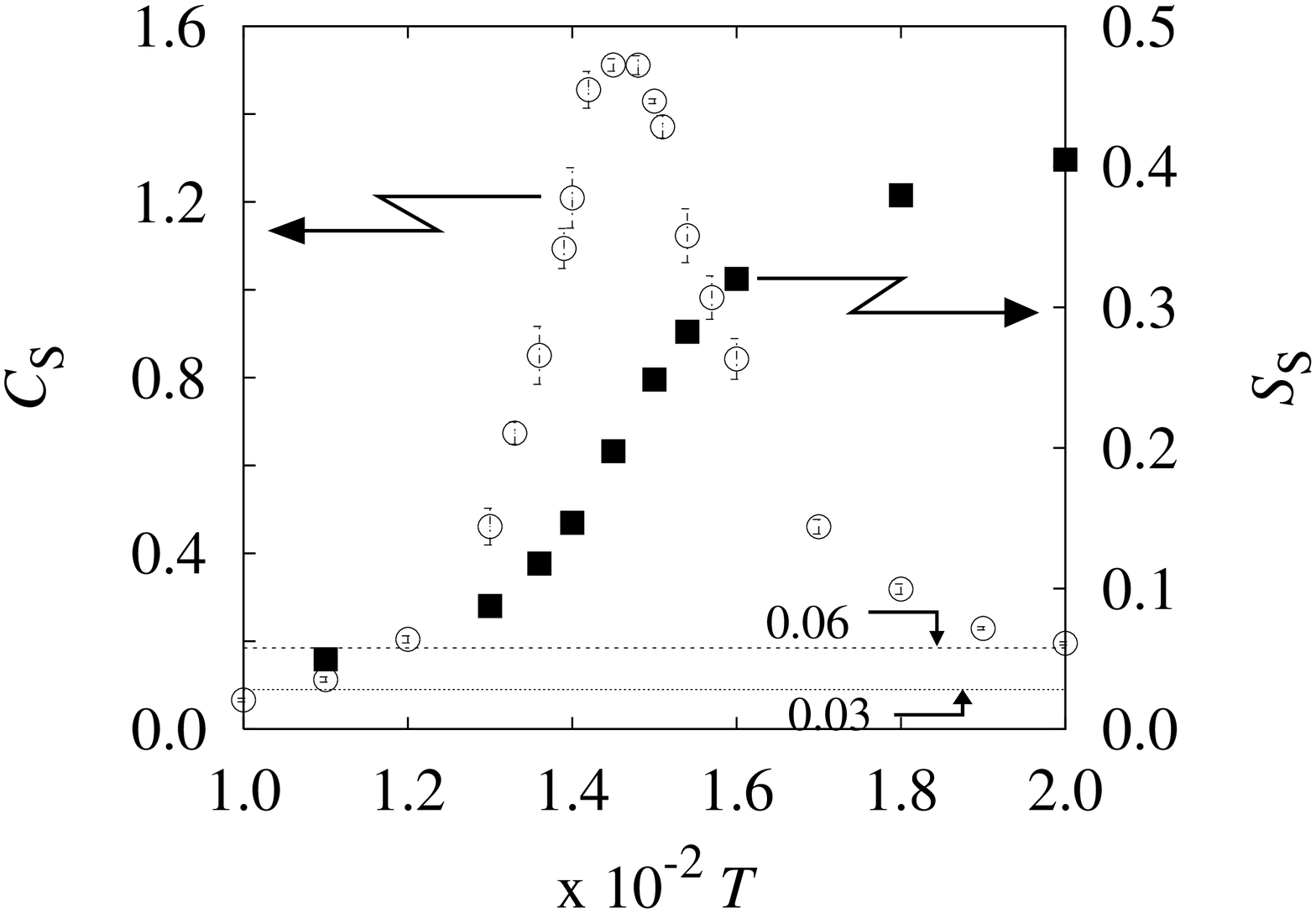}
\end{center}
\caption{Temperature dependences of the specific heat $C_{\rm S}$ and 
the entropy $S_{\rm S}$ of localized spins for the system with 
$6 \times 6$ unit cells. The upper and the lower dashed line represent 
$\frac{1}{12}\log2 \simeq 0.06$ and $0.34/12 \simeq 0.03$, respectively, 
where 0.34 is the value of the residual entropy of the AF Ising model 
on the triangular lattice. }
\label{fig:C_and_S_T0p01_0p05}
\end{figure}
Magnified views of $C_{\rm S}$ and $S_{\rm S}$ at the sharp peak 
are shown in Fig. \ref{fig:C_and_S_T0p01_0p05}.
From the results as in Fig. \ref{fig:C_and_S_T0p01_0p05}, it seems that 
$0.03 \lesssim S_{\rm S} \lesssim 0.06$ 
in the dodecamer phase. 
Thus, we may read off that 
there still remain degrees of freedom for localized spins 
even in the dodecamer phase. 
From this result, disordered psuedospin states which satisfy 
the AF interaction seem to appear in the dodecamer phase. 
\begin{figure}
\begin{center}
\includegraphics[width=6cm,keepaspectratio]{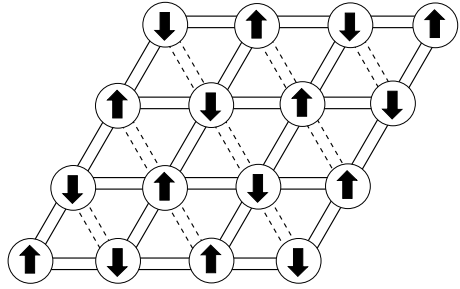}
\end{center}
\caption{The pseudo-AF phase. The uparrows and the downarrows represent 
the dodecamer with the out-spins on the tips and that with the in-spins, 
respectively. 
}
\label{fig:psdAFconf1}
\end{figure}
However, at further low temperatures, 
a pseudospin ordering may be expected to be realized 
considering the effective long-range nature due to the DE mechanism. 
A dodecamer phase without/with a long-range order of the pseudospins 
is called ``pseudo-para phase''/``pseudo-AF phase'', hereafter. 
The former case has only a translational symmetry breaking 
due to the dodecamer order, 
while the latter case also has a spin symmetry breaking 
due to some kind of pseudospin ordering 
in addition to the translational symmetry breaking. 
In Fig. \ref{fig:psdAFconf1}, we show a possible pseudo-AF phase 
as a candidate for the ground state. 
In the pseudo-AF phase, the strong bonds 
with the transfer integral $t_{1}$ linearly align in two directions, 
while the weak bonds with $t_{2}$ align in the other direction. 
The pseudo-AF phase has 
a rotational symmetry breaking in the lattice space 
in addition to the translational and the spin symmetry breaking. 
In the pseudo-AF phase, further kinetic energy gain may be expected 
because of the linear alignment of the strong bonds in two direction, 
that is, 
the motion of electrons on the pseudo-square lattice will be possible. 
As far as the DESI system with $4 \times 4$ and $8 \times 8$ unit cells 
has been investigated, our numerical calculations indicate that 
the pseudo-AF phase has the smallest energy in all dodecamer ordered states, 
i.e., the ground state of the DESI model 
on the kagom\'e lattice. 
According to these considerations, 
we expect that there is a possibility of a phase transition 
in the dodecamer ordered state, i.e., a transition 
from the pseudo-para phase to the pseudo-AF phase. 
However, the ground state in the thermodynamic limit 
cannot be clarified so far, 
and more detailed analysis at low temperatures 
for larger system sizes are needed.

\section{Discussion and Summary}

In the DESI system on the kagom\'e lattice, 
the long-range interaction due to the DE mechanism 
plays an important role to stabilize the dodecamer ordered state. 
In fact, a similar situation has been seen in the spin ice system
\cite{Bramwell_01_science}. 
The spin ice system has been discussed by two models: 
One has a ferromagnetic interaction due to the n.n. Ising interaction, 
and the other has that due to the long-range dipolar interaction.  
In particular, the latter model is called a dipolar spin ice model. 
According to MC calculations in the dipolar spin ice model, 
a double-peak structure in a specific heat has been observed 
as in the DESI system: 
A broad peak at higher temperature 
and a sharp peak at lower temperature\cite{Melko_01}. 
The sharp peak corresponds to the spin cluster ordered state. 
Such a cluster ordered state 
is not observed in the spin ice model with the n.n. Ising interaction.
These results indicate 
the long-range nature of the interaction  
plays an important role to create a cluster ordering 
even in the frustrated spin systems as well as the electron systems. 
Although the spin ice system is an insulator, 
the DESI system is considered to be metallic.
In this way, a cluster ordering might be generic to frustrated systems. 
In general, the frustration suppresses the usual ordered states and 
leads to the degenerate states at low temperatures. 
In the frustrated systems, 
the residual interaction selects the cluster ordered state 
among the macroscopic degenerate states.

\begin{figure}
\begin{center}
\includegraphics[width=8cm,keepaspectratio]{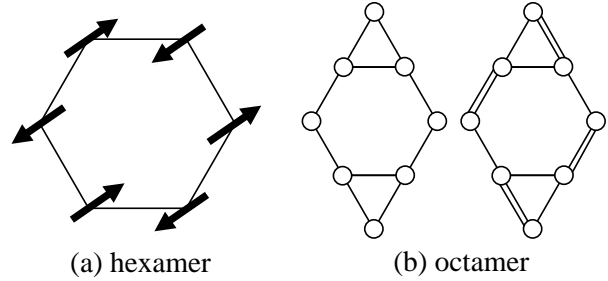}
\end{center}
\caption{(a) Hexamer in ${\rm ZnCr_{2}O_{4}}$. 
The $S=3/2$ localized spins 
align antiferromagnetically in an arbitrary direction.
(b) Octamers in ${\rm CuIr_{2}O_{4}}$: 
(left) Octamer composed of eight ${\rm Ir^{3+}}$ ($S=0$) 
which is represented by open circles. 
(right) Octamer composed of eight ${\rm Ir^{4+}}$ ($S=1/2$) 
which is represented by open circles, 
where the double lines represent the dimer-singlets of ${\rm Ir^{4+}}$.}
\label{fig:hexa_octa}
\end{figure}

Let us consider a possibility of 
the cluster order in realistic systems with the frustration. 
As mentioned in the introduction, 
there are several examples of a cluster order in nature: 
the hexamer in ${\rm ZnCr_{2}O_{4}}$\cite{Lee_02} 
[see Fig. \ref{fig:hexa_octa}(a)],
the octamer in ${\rm CuIr_{2}S_{4}}$\cite{Radaelli} 
[see Fig. \ref{fig:hexa_octa}(b)]
and so forth. 
In the cubic spinel ${\rm ZnCr_{2}O_{4}}$, 
${\rm Cr^{3+}}$ ($S=3/2$) resides on the pyrochlore structure
and interact with each other antiferromagnetically\cite{Lee_00}. 
At low temperatures, 
a hexagonal AF spin cluster formation is realized and 
the pyrochlore sites are tiled by the hexamers\cite{Lee_02}. 
In a thiospinel ${\rm CuIr_{2}S_{4}}$, 
mixed-valence states of ${\rm Ir^{3+}}$ ($S=0$)
and ${\rm Ir^{4+}}$ ($S=1/2$)
are realized in the insulating phase at low temperatures 
and they occupy octahedral sites
\cite{Oda,Kumagai,Matsuno}. 
In this case, 
the system have a charge frustration as well as a spin frustration. 
In the insulating phase, Ir sites are tiled by two kinds of octamers:
One consists of eight ${\rm Ir^{3+}}$ 
and the other consists of eight ${\rm Ir^{4+}}$ 
with four dimer-singlets. \cite{Radaelli}
In this way, the cluster order 
can be realized in realistic frustrated systems. 
Such a cluster order may be explained by including the effect of 
the residual interactions which are neglected in the non-frustrated system.

In conclusion, there are at least two steps for the entropy release 
in the DESI system on the kagom\'e lattice, 
which is indicated by the peak structures in the specific heat. 
The ice state appears at intermediate temperatures 
due to the n.n. ferromagnetic interaction. 
From the estimated entropy, 
it has been clarified that 
there is a large number of thermally degenerate states 
in the ice state. 
As the temperature is lowered, 
a large part of the entropy in the system 
is released by the dodecamer ordering 
with the translational symmetry breaking. 
The dodecamer order is driven by the kinetic energy gain and the frustration: 
The dodecamer is the kinetic-energetically stable path for electrons. 
Furthermore, a possibility of a transition in the dodecamer phase 
is indicated by our MC results: 
(1) The estimated entropy $S_{\rm S}$ seems to 
remain in the dodecamer ordered phase 
because of the existence of two kinds of dodecamers 
which can be regarded as the pseudospins. 
(2) The pseudo-AF phase is the ground state in the system 
with small-size clusters. 
These result possibly indicate that 
there is a transition from the pseudo-para phase to the pseudo-AF phase
to gain the further kinetic energy. 
It is expected that the entropy in the system will be completely released by 
the ordering of the pseudospins, i.e., two kinds of dodecamers. 
However, more detailed analysis are needed to clarify this point.

\section*{Acknowledgements}

We  acknowledge M. Miyazaki for sincerely reading of our paper. 
We would like to thank 
S. Katsurada and Y. Shinohara for computational supports.
The numerical computations have been performed 
mainly using the facilities of the super AOYAMA+ project (SAPP). 
This work was partially supported by a Grant-in-Aid for 21st COE program 
from the Ministry of Education, Culture, 
Sports, Science and Technology of Japan.

\appendix
\section{Monte Carlo Method}

Here, we explain the application of the MC method 
for the Hamiltonian (\ref{eq:DESImodel}) in detail\cite{Yunoki_98}. 
We define abbreviations $\{ c_{i},c_{i}^{\dagger} \} \equiv 
( c_{1},\cdots ,c_{N_{\rm S}},
c_{1}^{\dagger}, \cdots ,c_{N_{\rm S}}^{\dagger} )$ and 
$\{ \tau_{i} \} \equiv ( \tau_{1},\cdots ,\tau_{N_{\rm S}})$ 
which represent degrees of freedom for electrons 
and localized spins (adiabatic fields), respectively. 
Using Eq. (\ref{eq:DESImodel}), 
the partition function of the system in the grand canonical 
ensemble is written as 
\begin{align}
Z &= {\rm Tr_{C} Tr_{F}}
\exp \left[ - \beta \left \{ 
\sum_{\langle i,j \rangle} t \left( \tau_{i}, \tau_{j} \right) 
\left( c_{i}^{\dagger} c_{j} + h.c.\right)
- \mu \sum_{i} c_{i}^{\dagger}c_{i} \right \} \right], \\
&= {\rm Tr_{C} Tr_{F}}
\exp \left[ - \beta \left(  \hat{H}_{\rm K} 
- \mu \sum_{i} c_{i}^{\dagger}c_{i} \right) \right], 
\label{eq:partition_fnc}
\end{align}
where ${\rm Tr_F}$ and ${\rm Tr_C}$ represent traces over 
$\{ c_{i},c_{i}^{\dagger} \}$ and $\{ \tau_{i} \}$, respectively. 
And, $\mu$ represents the chemical potential. 
Here, we define the kinetic energy part of the Hamiltonian as 
$\hat{H}_{\rm K} \equiv \sum_{\langle i,j \rangle} 
t \left( \tau_{i}, \tau_{j} \right) 
\left( c_{i}^{\dagger} c_{j} + h.c.\right)$. 
Once a localized spin configuration $\{ \tau_{i}\}$ is given, 
it is easy to diagonalize the Hamiltonian (\ref{eq:DESImodel}) 
because it is represented 
by the quadratic form of fermion creation and annihilation operators, 
i.e., $N_{\rm S} \times N_{\rm S}$ matrix: 
$\hat{H}_{\rm K} = \sum_{\langle i,j \rangle}c_{i}^{\dagger}H_{ij}c_{j}$, 
where $H_{ij} \equiv t(\tau_{i},\tau_{j})$ 
for n.n. pairs of $i$ and $j$, otherwise 0. 
Thus, the trace operation ${\rm Tr_{F}}$ is 
easily performed through the energy eigenvalues 
$\varepsilon_{k}(\{ \tau_{i}\})$ of $\hat{H}_{\rm K}$ 
for a given configuration $\{ \tau_{i}\}$. 
Finally, the partition function (\ref{eq:partition_fnc}) 
can be written as 
\begin{equation}
Z = {\rm Tr_C} \exp \left[ -\beta F_{\rm eff}( \{ \tau_{i} \} ) \right], 
\label{eq:eff_partition_fnc}
\end{equation}
with the effective free energy
\begin{equation}
F_{\rm eff} (\{ \tau_{i} \} ) = - k_{\rm B} T 
\sum_{k=1}^{N_{\rm S}} \log \left( 
1+ \exp \left\{ - \beta \left[ \varepsilon_{k}( \{ \tau_{i} \} )-\mu  \right] \right \} \right).
\label{eq:Feff}
\end{equation}
Probability distribution of a certain configuration $\{ \tau_{i}\}$ 
is given by
\begin{equation}
P( \{ \tau_{i}\} ) =
\exp \left[ -\beta F_{\rm eff}( \{ \tau_{i} \} ) \right]/Z, 
\label{eq:weight}
\end{equation}
whose numerator is regarded as the Boltzmann factor.

Thermal average of arbitrary observable 
$\hat{A} (\{c_{i},c_{i}^{\dagger}\},\{ \tau_{i} \}) $  
is given as follows in the grand canonical ensemble: 
\begin{align}
\langle \hat{A}(\{c_{i},c_{i}^{\dagger}\},\{ \tau_{i} \}) \rangle 
&\equiv 
\frac{{\rm Tr_C Tr_F} 
\left[ \hat{A}(\{c_{i},c_{i}^{\dagger}\},\{ \tau_{i} \}) 
{\rm e}^{-\beta \hat{H}} \right]}
{{\rm Tr_C Tr_F} \left[ {\rm e}^{-\beta \hat{H}} \right] }.
\label{eq:thermal_ave_gc}
\end{align}
Using Eq. (\ref{eq:weight}), the thermal average is also written as
\begin{align}
\langle \hat{A}(\{c_{i},c_{i}^{\dagger}\},\{ \tau_{i}\}) \rangle &= 
\langle \langle 
\hat{A}(\{c_{i},c_{i}^{\dagger}\},\{ \tau_{i}\}) 
\rangle_{\rm F} \rangle_{\rm C},
\label{eq:thermal_ave_tot}
\end{align}
where  $\langle \cdots \rangle_{\rm F}$ represents the grand canonical average 
for a fixed configuration $\{ \tau_{i}\}$ defined by 
\begin{align}
\langle \cdots \rangle_{\rm F} &\equiv 
\frac{{\rm Tr_F} 
\left[ \cdots
{\rm e}^{-\beta \hat{H} }\right] }
{{\rm \rm Tr_{F}} 
\left[ {\rm e}^{-\beta \hat{H}} \right] },
\label{eq:thermal_ave_f}
\end{align}
while $\langle \cdots \rangle_{\rm C}$ is the thermal average over 
configurations of $\{ \tau_{i}\}$ defined by
\begin{align}
\langle \cdots \rangle_{\rm C} &\equiv 
{\rm Tr_{C}}\left [  
\cdots P(\{ \tau_{i}\})
\right ].
\label{eq:thermal_ave_c}
\end{align}
Note that, in the thermal average for observables 
without the electron operator $ \{ c_{i},c_{i}^{\dagger} \}$, 
e.g., $v_{i}^{z}$ (Eq. (\ref{eq:chiral})) 
and $d_{i}$ (Eq. (\ref{eq:order_param})), 
$\langle \cdots \rangle$ becomes $\langle \cdots \rangle_{\rm C}$.

\begin{figure}
\begin{center}
\includegraphics[width=8cm,keepaspectratio]{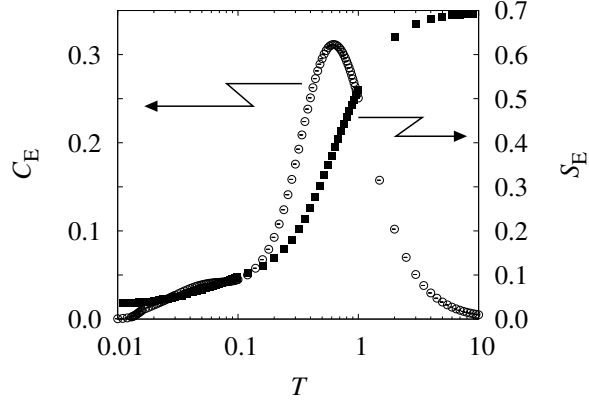}
\end{center}
\caption{Temperature dependence of $C_{\rm E}$ and $S_{\rm E}$. 
$C_{E}$ has a peak at $T \simeq 0.6$, 
which is almost the same order of $t_{1}$ and $t_{2}$.}
\label{fig:Ce_Se}
\end{figure}

The specific heat is defined as a fluctuation 
of the energy in the form
\begin{equation}
C(T) = \frac{ \langle \hat{H}^{2} \rangle -\langle \hat{H} \rangle^{2} }
{T^{2}}.
\label{eq:spe}
\end{equation}
$C(T)$ can be decomposed into 
electronic part $C_{\rm E}(T)$ 
and spin part $C_{\rm S}(T)$
as follows: 
\begin{align}
C(T) &= C_{\rm E}(T) + C_{\rm S}(T),
\end{align}
where 
\begin{align}
C_{\rm E}(T) & \equiv \frac{1}{T^{2}}
\langle 
E_{2}-E_{1}^{2}
\rangle_{\rm C},
\label{eq:spe_ele}
\end{align}
\begin{align}
C_{\rm S}(T) & \equiv \frac{1}{T^{2}}
\left(
\langle 
E_{1}^{2}
\rangle_{\rm C} 
- 
\langle 
E_{1}
\rangle_{\rm C}^{2} 
\right). 
\label{eq:spe_fld}
\end{align}
Here, we set $E_{1} \equiv \langle \hat{H} \rangle_{\rm F}$ and 
$E_{2} \equiv \langle \hat{H}^{2} \rangle_{\rm F}$. 
Eq. (\ref{eq:spe_ele}) is given by the electronic energy fluctuation 
averaged over configurations $\{ \tau_{i} \}$, 
while Eq. (\ref{eq:spe_fld}) is given by the energy fluctuations 
among spin configurations $\{ \tau_{i} \}$. 
Thus, it is easily found that 
Eq. (\ref{eq:spe_ele}) does not include fluctuation of localized spins 
while Eq. (\ref{eq:spe_fld}) does not include fluctuation of electrons. 
$E_{1}$ and $E_{2}$ are calculated as 
\begin{align}
E_{1}
&= \sum_{k} \varepsilon_{k}(\{ \tau_{i}\})
f_{\rm F}(\varepsilon_{k}(\{ \tau_{i}\})), \\
E_{2}
&= \sum_{k} \varepsilon_{k}(\{ \tau_{i}\})^{2}
f_{\rm F}(\varepsilon_{k}(\{ \tau_{i}\})) \nonumber \\
&+
\sum_{k \neq l} \varepsilon_{k}(\{ \tau_{i}\})\varepsilon_{l}(\{ \tau_{i}\})
f_{\rm F}(\varepsilon_{k}(\{ \tau_{i}\}))
f_{\rm F}(\varepsilon_{l}(\{ \tau_{i}\})).
\end{align}
where $f_{\rm F}(x)$ is the Fermi distribution function given in the form, 
\begin{align}
f_{\rm F}(x)=\frac{1}{{\rm e}^{\beta (x-\mu)}+1}.
\label{eq:fermi}
\end{align}

In the grand canonical ensemble, 
the entropy of electronic part per site in the DESI system, 
$S_{\rm E}(T=\infty)$ is $\log2$ as follows: 
\begin{align}
S_{\rm E}(T=\infty)
&= \int_{0}^{\infty}dT \frac{C_{\rm E}(T)}{T} \nonumber \\
&= \left \langle 
\int_{0}^{\infty} dT \frac{1}{T^{3}}
(E_{2} - E_{1}^{2})
\right \rangle_{\rm C} \nonumber \\
&= \langle \log2 \rangle_{\rm C} \nonumber \\
&= \log2.
\end{align}
In the DESI system, 
the entropy of localized spin is $\log2$, 
since each localized spin has two degrees of freedom. 
Therefore, total entropy per site in the DESI system is $\log4$.
Temperature dependences of $C_{\rm E}$ and $S_{\rm E}$ for the system 
with 6 $\times$ 6 unit cells are shown in Fig. \ref{fig:Ce_Se}. 
A peak can be found at around $T = t_{1} \sim t_{2}$ 
in Fig. \ref{fig:Ce_Se}. 
In the electronic part of the specific heat, $C_{\rm E}$, 
any remarkable structures are not observed 
at sufficiently low temperatures $T \sim 0.01$ 
where the dodecamer order is realized.

\end{document}